\documentclass[%
 reprint,
 amsmath,amssymb,
 aps,onecolumn,showkeys
]{revtex4-2}

\usepackage{graphicx}
\usepackage{dcolumn}
\usepackage{bm}

\usepackage{xcolor}

\begin{document}


\title{Diffusion of Tracer Particles in Early Growing Biofilms\\A Computer Simulation Study}

\author{Fabi\'an A. Garc\'ia Daza$^{1}$
\footnote{Current address: Euskal Oxcitas Biotek SL, 48008, Bilbao, Spain}}%
	
\author{\'Alvaro Rodr\'iguez-Rivas$^{3}$}

\author{Fernando Govantes$^{4,5}$}

\author{Alejandro Cuetos$^{1,2}$}
\email{acuemen@upo.es}

\affiliation{$^1$Department of Physical, Chemical and Natural Systems, Pablo de Olavide University, Seville, Spain}
\affiliation{$^2$ Center for Nanoscience and Sustainable Technologies (CNATS), Pablo de Olavide University, Seville, Spain}

\affiliation{$^3$ Departamento de F\'isica Aplicada I, Escuela Polit\'ecnica Superior, Universidad de Sevilla, 41011, Seville, Spain}


\affiliation{$^4$ Centro Andaluz de Biolog\'{\i}a del Desarrollo (Universidad Pablo de Olavide, Consejo Superior de Investigaciones Cient\'{\i}ficas y Junta de Andaluc\'{\i}a), Seville, Spain}

\affiliation{$^5$ Departamento de Biolog\'ia Molecular e Ingenier\'ia Bioqu\'imica, Universidad Pablo de Olavide, Seville, Spain}

\begin{abstract}
	
The diffusion of particles in complex media has gained significant interest due to its dual relevance: probing the viscoelastic properties of materials via microrheology and assessing the extent of particle displacement over time. In this work, we explore the early-stage diffusion of a tracer particle within a developing bacterial biofilm using implicit-solvent Brownian dynamics simulations. At these initial stages, bacterial colonies form two-dimensional structures that expand through cell growth and division. Employing an agent-based computational model (IbM), we analyse the passive diffusion of a spherical tracer within colonies of varying compaction levels. Our findings reveal that, at very short timescales, tracer diffusion follows a standard diffusive regime, modulated by colony ageing. However, at longer times, the dominant factor governing tracer motion is colony growth, which effectively confines the tracer within the expanding structure, except in cases where the microcolony is highly unstructured or the tracer is sufficiently small. Additionally, through MR techniques, we quantify the elastic and viscous moduli of the growing microcolony, offering insight into its evolving viscoelastic behavior.
\end{abstract}

\keywords{Tracer, diffusion, biofilm, microrheology, computer simulation, IbM model}
\maketitle


\section{Introduction}

The study of single-particle diffusion, often referred to as tracer particle diffusion, in complex fluid systems with internal structure, such as liquid crystals, polymers, biological media, and other specialized materials, has emerged as a topic of broad interdisciplinary interest. This interest arises from the ability to infer physical properties of the host medium by analyzing the transport behavior of the tracer particle, sometimes at the nanometer scale, even in the presence of impurities or nutrients \cite{SAL01,ZHA10,COP23}. Notably, it has been shown that the transport properties of a tracer particle can reveal structural and viscoelastic characteristics of the surrounding medium \cite{OUE10,JAR22}, due to local distortions in the fluid caused by the tracer particle. This interplay between the tracer and the host fluid underpins the methodologies grouped under microrheology (MR) \cite{MAS96,PUE14}. MR techniques, classified on passive MR if the motion of the tracer is only induced by thermal fluctuations \cite{MAN95,MAN00}, and active MR if an external force is applied \cite{SQU05,PUE14,KUH15}, have emerged as alternatives to the determination of rheological properties using macroscopic techniques. These methodologies, as described in the literature, have become viable alternatives to macroscopic methods for characterizing rheological properties and have been applied experimentally and computationally in systems such as liquid crystals \cite{DAZ22b,DAZ22,DAZ23,DAZ24}, polymer and colloidal suspensions \cite{AMB96,CRI10,MED18}, and materials of biological or pharmaceutical relevance \cite{FEN01,WAT14,HAR19,SIN23}.

Tracer particle diffusion plays a vital role in growing cell communities, such as tissues, tumors, cell colonies, and bacterial biofilms. In these systems, even passive tracer particles experience altered dynamics due to the active nature of cell growth and division, which involves nutrient, energy, and momentum exchange with the environment. As a result, a dynamics distinct from thermal fluctuations emerges, often manifesting in super-diffusive behavior, as reported in cancer multicellular spheroids \cite{SIN23}, animal cells \cite{WEI06}, and bacterial biofilms \cite{ROG08}. Theoretical studies have focused on how the growth of the medium affects the dynamics of the tracer, and less on the modification of the dynamics by the individual growth of each cell \cite{YUS16,AB20,CHE23}. In this context, tracer diffusion also provides a window into the rheological properties of biological materials, as MR has been applied to measure viscoelastic properties in cellular assemblies \cite{WEI06,WIR09,BIR14,HAR19}. From a biological perspective, diffusion processes are inherently significant, particularly in cellular communities, where particle diffusion influences cellular interactions. Extensive research has explored these processes, emphasizing their critical roles in tissues, tumors, bacterial biofilms, and other cellular communities \cite{KO20,FUL19,LAH13,YUA22}. This study aims to improve the understanding of particle diffusion in cellular systems. The findings presented here are expected to have relevance across various biological scenarios, including bacterial biofilm growth, tissue development, and tumor formation. Furthermore, the insights may extend to materials science, particularly in systems where complex structures emerge due to polar growth, such as the formation of calcium-silicate-hydrate (CSH) particles during cement hardening \cite{RIC99,ROL09} or the growth of TiO$_2$ nanowires \cite{BER13}. To this end, we performed computer simulations to examine the diffusion behavior of tracer particles in early biofilm stages. At these early stages, biofilms are predominantly two-dimensional, transitioning to three-dimensional structures as they develop \cite{GRA14,PIN12}. Notably, certain experimental setups have shown biofilms confined between two surfaces, leading to a persistent two-dimensional structure \cite{VOL08}. 

In this study, we employ an agent-based computational model (IbM) for simulating biofilm growth, developed by our group and previously applied to both two- and three-dimensional biofilms \cite{ACE18,LOB21,DEL22,LOB24}.  These prior works revealed that the biofilm's compactness and structural characteristics are shaped by the balance between bacterial diffusion and growth dynamics. Compact colonies arise when growth and division dominate the system. By contrast, when diffusion plays the predominant role, cells spread widely over the surface, leading to open and less compact structures. These two modes of growth are termed closed and open growth, respectively \cite{ACE18}.

The goals of this study are twofold. First, we investigate how cell growth and division influence tracer particle diffusion, paying special attention to the differences that arise in closed versus open growth scenarios. The active nature of cell growth and division suggests inherent differences between tracer diffusion in biofilms and in other fluids containing elongated particles, such as liquid crystals \cite{DAZ22,DAZ24}. To investigate these differences, we have analyzed key observables commonly used in MR and the study of transport phenomena in complex fluids. The results provide valuable insights into microrheology applications for systems with self-replicating particles, such as biofilms and other cell-based communities. Second, we present findings on the passive diffusion of nanoparticles escaping from a growing biofilm, a result of biological relevance for understanding and quantifying nanoparticle extrusion in bacterial populations.

\section{Methodology}\label{sec_meth}
In this work, we investigate the diffusion behavior of a spherical tracer within a growing bacterial microcolony. Our focus is on the initial stages of microcolony growth, where the system can be approximated as two-dimensional. Our model \cite{ACE18,LOB21,DEL22} does not consider active particle motion, with the tracer particle and bacteria being affected by inter-particle interactions and passive diffusion. Bacteria are represented as a bidimensional projection of spherocylinders, formed by a cylinder with an initial length denoted as $L_0$ and two hemispherical ends with a diameter of $\sigma$. In this study, we replicated key attributes of \textit{Pseudomonas putida}, such as its length-to-diameter ratio ($L_0^{*} = L_0/\sigma+1= 2.6$), which has been previously documented in experimental studies and incorporated into recent computational research \cite{PAL15,DEL22}. Over time, bacterial growth is modelled by increasing their lengths according to $L(t)=L_0\exp(r^m(t-t^m_a))$, where $L_0$ represents the initial particle length, $t_a$ the time of the previous division and $r^m$ is the growth rate of bacterium $m$. The growth rate is randomly selected from a Gaussian distribution centred around a constant rate $r$, with a relative standard deviation of $s/r=0.1$. This stochastic variability is consistent with observations, and has been found in experimental systems \cite{FAC19,SI19}.  Upon reaching a maximum elongation of $L_m^{*}=2L_0^{*}$, each bacterium undergoes division, resulting in two particles with identical $L_0^{*}$ ratios \cite{TAH15}. These new particles continue to grow according to the same growth model. By contrast, the tracer particle is represented as a sphere with a fixed diameter $d_t$, ranging from $10^{-3}\sigma$ to $\sigma$. 

Interactions between particles are mediated via a truncated and shifted Kihara potential 
\begin{equation}
  \label{eq1}
  U_{ij}=
  \begin{cases}
    4\epsilon_{ij} \left[\left(\frac{\sigma_{ij}}{d_m}\right)^{12}-\left(\frac{\sigma_{ij}}{d_m}\right)^{6}+\frac{1}{4}\right],& \text{if } d_m\leq \sqrt[6]{2}\sigma_{ij}\\
    0,& \text{if } d_m> \sqrt[6]{2}\sigma_{ij},
  \end{cases}
\end{equation}
where the interaction between pairs of particles $i$ and $j$ is described by the interaction strength $\epsilon_{ij}$. In this work we have set $\epsilon_{ij}= k_B T$ for rod-rod , and $\epsilon_{ij}= 10k_B T$ for tracer-rod interactions, respectively, where $k_B$ is the Boltzmann constant and $T$ the temperature. The sum of the particles' radii is represented by $\sigma_{ij}$, and the minimum distance between them, calculated in the rod-rod case using the algorithm proposed by Vega and Lago \cite{VEGA199455}, is denoted as $d_m$. The potential energy term $U_{ij}$ exclusively accounts for steric interactions between particles and neglects any attractive interactions.

The movement of particles in the system is simulated using Brownian dynamics (BD) simulations at a constant volume and temperature. In BD simulations, the Langevin equation is employed to integrate the positions and orientations of the particles over time, generating their trajectories \cite{LOWEN1994}. Specifically, for rod-like particles, the position of the center of mass $\mathbf{r}_j$ and the orientation $\hat{\mathbf{u}}_j$ of each particle are updated over time using the following set of equations: 
\begin{align}
    \mathbf{r}_j^\parallel(t+\Delta t) = & \mathbf{r}_j^\parallel(t) + \frac{D_\parallel}{k_{\text{B}}T}\mathbf{F}_j^\parallel\Delta t + \sqrt{2D_\parallel\Delta t}R^\parallel\hat{\mathbf{u}}_j^\parallel(t)\\
    \mathbf{r}_j^\perp(t+\Delta t) =&\mathbf{r}_j^\perp(t)  + \frac{D_\perp}{k_{\text{B}}T}\mathbf{F}_j^\perp\Delta t\nonumber\\ 
    &+ \sqrt{2D_\perp\Delta t}\left[R_1^\perp\hat{\mathbf{v}}_{j,1}^\perp(t) + R_2^\perp\hat{\mathbf{v}}_{j,2}^\perp(t)\right]\\
    \hat{\mathbf{u}}_j(t+\Delta t) = &\hat{\mathbf{u}}_j(t) + \frac{D_\vartheta}{k_{\text{B}}T}\mathbf{T}_j \times \hat{\mathbf{u}}_j(t) \Delta t\nonumber\\
    &+ \sqrt{2D_\vartheta\Delta t}\left[R_1^\vartheta\hat{\mathbf{w}}_{j,1}^\perp(t) + R_2^\vartheta\hat{\mathbf{w}}_{j,2}^\perp(t)\right],    
\end{align}\label{eq2}
where $\mathbf{r}_j^\parallel$ and $\mathbf{r}_j^\perp$ represent the projections of the position of particle $j$ in the parallel and perpendicular directions to $\mathbf{u}_j$, respectively. The forces acting on the particle are split into parallel ($\mathbf{F}_j^\parallel$) and perpendicular ($\mathbf{F}_j^\perp$) components, while $\mathbf{T}_j$ represents the torque acting on the particle, easily obtained from the interaction potential \cite{VEG90}. The values $R^\parallel$, $R_1^\perp$, $R_2^\perp$, $R_1^\vartheta$, and $R_2^\vartheta$ are independent Gaussian random numbers with a mean of zero and variance 1. The unit vectors $\hat{\mathbf{v}}_{j,m}$ and $\hat{\mathbf{w}}_{j,m}$ (where $m=1,\:2$) are perpendicular to $\hat{\mathbf{u}}_j$. To determine the translational and rotational diffusion coefficients ($D_{\parallel}$, $D_{\perp}$, and $D_{\vartheta}$), we utilize the numerical solution of integral expressions obtained from the induced-force method for uniaxial particles proposed by Bonet Avalos and colleagues \cite{BONETAVALOS1994193}. It is important to note that these diffusion coefficients depend on the size of the bacterium and are calculated for each bacterium at every point in time. The specific expressions for these diffusion coefficients, given a particular aspect ratio $L^{*}=L(t)/\sigma+1$, are as follows:
\begin{align}
    \frac{D_\parallel}{D_0} = & -0.0198\ln(L^{*}) + 0.0777 +\frac{0.0437}{L^{*}} - \frac{0.0158}{(L^{*})^2},\\
    \frac{D_\perp}{D_0} = & -0.0119\ln(L^{*}) + 0.0452 +\frac{0.0796}{L^{*}} - \frac{0.0190}{(L^{*})^2},\\
    \frac{D_\vartheta\sigma^2}{D_0} = & -0.0002\ln(L^{*}) + 0.0012 - \frac{0.0243}{L^{*}} + \frac{0.3233}{(L^{*})^2} \nonumber\\
    &  + \frac{0.2597}{(L^{*})^3} - \frac{0.0483}{(L^{*})^4}, 
\end{align}
where $D_0=\sigma^2/\tau$, with $\tau=\sigma^2/\kappa_B\nu$ the unit  ot time, being $\kappa_B$ the Boltzmamn constant, and $\nu$ the viscosity.

\begin{figure*}[!t]
	\includegraphics[scale=0.9]{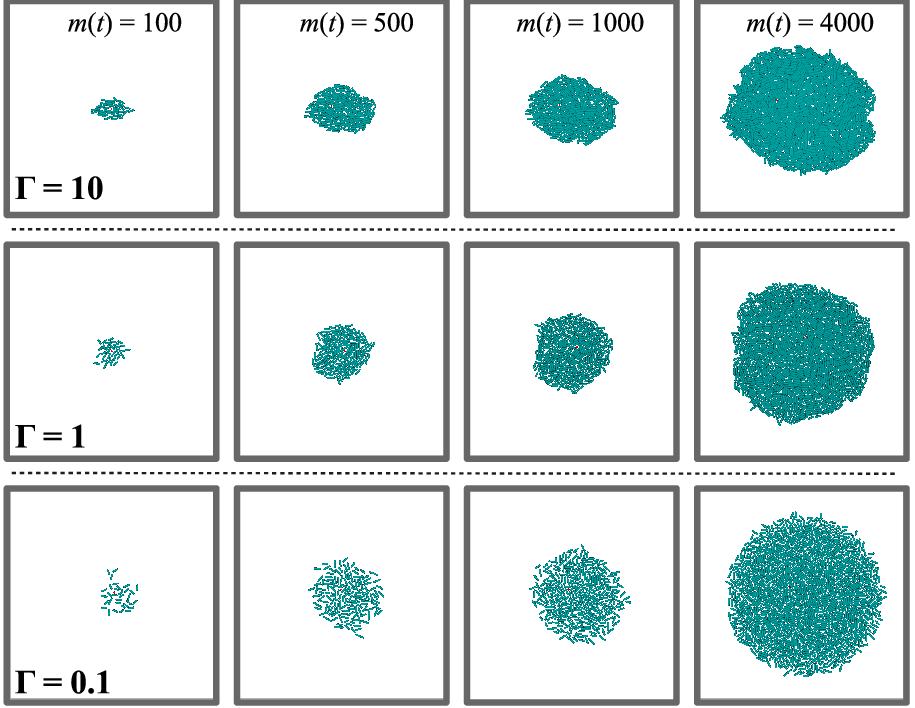}
	\caption{Representative snapshots of a tracer with a diameter of $\sigma$ diffusing within an expanding bacterial system at various biomass levels are shown. The progression of the bacterial colony, marked by an increase in biomass $m(t)$, is displayed from left to right. The different growth rates values explored in this work, spanning from $\Gamma=10$ to $\Gamma=0.1$, are depicted from top to bottom.}\label{fig1}
\end{figure*}

For tracer particles, the position $\mathbf{r}_t$ of a spherical particle changes in time as
\begin{equation}\label{eq8}
    \mathbf{r}_t(t+\Delta t) = \mathbf{r}_t(t) + \frac{D_t}{k_{\rm{B}}T}\mathbf{F}_t(t)\Delta t + \sqrt{2D_t\Delta t} \mathbf{R}_0(t),
\end{equation}
where $D_t=D_0/(3\pi d_t)$ is the tracer diffusion coefficient at infinite dilution, $\mathbf{F}_t$ represents the total force exerted by other particles on the tracer, and $\mathbf{R}_0=\{R_x,R_y,R_z\}$ is a Gaussian random vector with unit variance and zero mean. The time step in our $\Delta t$ simulations was limited to values less than $10^{-3}\tau$, ensuring that at each step the displacement of any degree of freedom of the centre of mass of each particle (tracer or bacterium) was less than $0.01\sigma$.

In all the scenarios investigated, the initial setup involved two bacteria with an elongation of $L_0^{*}$ positioned in parallel to each other. They were separated by a distance of $2\sigma$ from their respective geometrical centres. Additionally, we placed one tracer particle between the centres of the bacteria. This spacing provides physical separation while reflecting typical bacterial dimensions. We verified the robustness of our results by testing alternative seeding geometries (e.g., square configurations with four bacteria) and found negligible differences in tracer dynamics, even under strong bacteria growth conditions. To prevent the tracer particle from escaping the microcolony, its position was fixed, while the bacteria were allowed to grow until reaching a predefined biomass threshold ($m_{\rm{start}}$). The amount of biomass in the microcolony is defined as 
\begin{equation}
    m(t)=\sum_{i=1}^{N(t)} L_i^{*}(t),   
\end{equation}
where $N(t)$ represents the number of bacteria at time $t$. It should be noted that both $N(t)$ and $L_i^{*}(t)$ are time-dependent, which consequently affects the biomass $m(t)$ over time. Once the biomass reaches a threshold of $m_{\rm{start}}=100$, the tracer particle is permitted to move due to the thermal noise and the interaction with the surrounding bacteria, undergoing temporal evolution. We highlight that the tracer's dynamics are expected to be affected by the value of $m_{start}$, as it experiences different pressure environments depending on the colony’s growth stage. While this is an important consideration, a thorough analysis on the influence of $m_{start}$ on the tracer’s dynamics is beyond the scope of this work. Using the methodology described above, $1000$ independent trajectories have been simulated for each case, up to a limit of $m_{\rm{max}}=4000$ per trajectory.

\section{Results}

In this section, we examine the interplay between bacterial colony geometry, determined by the combined effects of bacterial diffusion and growth, and the dynamic response of a spherical tracer, with a size comparable to the thickness of a single bacterial cell.  The different scenarios for biofilm growing are characterized by the parameter $\Gamma$ \cite{ACE18,DEL22}, defined as:

\begin{equation}
	\Gamma=t_{\rm{dif}}/t_{\rm{gr}},
\end{equation}  
\noindent where $t_{\rm{dif}}$ is the time needed by an isolated bacterial cell with constant $L^*=L^*_0$ to diffuse a distance $\sigma$ by Brownian motion, while $t_{\rm{gr}}$ is the time required by an average cell to reach the division size $L^*_m$ from $L^*_0$.  Previous studies have shown that $\Gamma$ is a suitable indicator of the regime in which the biofilm evolves \cite{ACE18,DEL22}. 

\begin{figure} [!t]
	\includegraphics{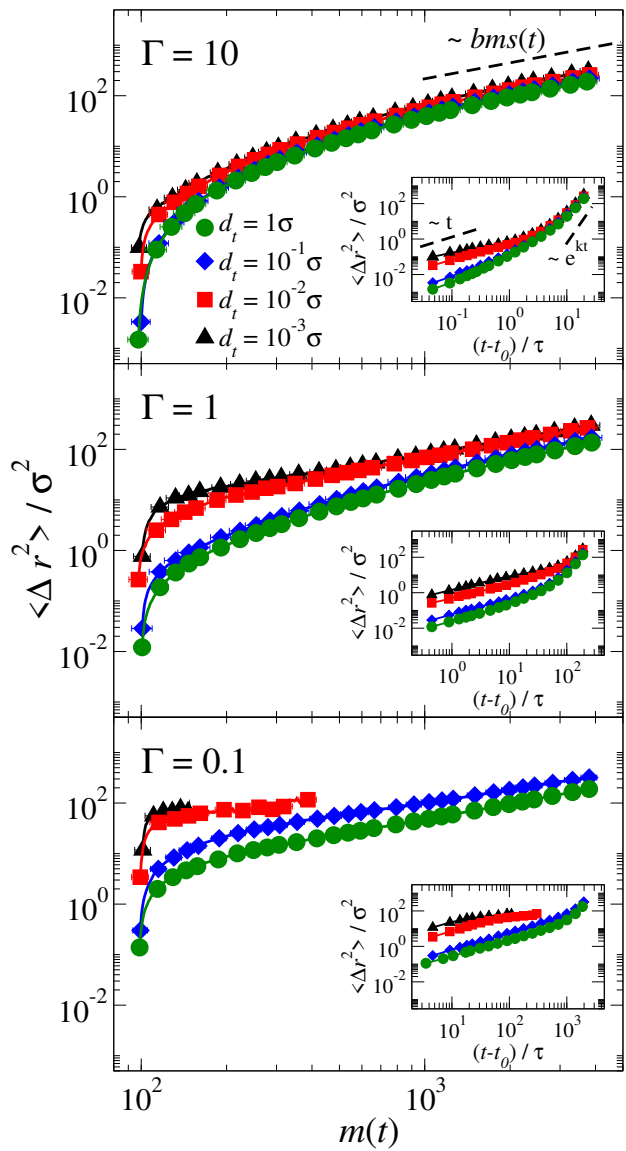}
	\caption{The mean-squared displacements (MSDs) of tracers with varying sizes $d_t=1\sigma$ (circles), $d_t=10^{-1}\sigma$ (diamonds), $d_t=10^{-2}\sigma$ (squares), $d_t=10^{-3}\sigma$ (triangles), are reported in relation to the biomass of the bacterial colony. The inset figure depicts MSD plotted against time. Three distinct values of bacterial growth rates are depicted: $\Gamma=10$ (upper panel), $\Gamma=1$ (middle panel), and $\Gamma=0.1$ (lower panel).}\label{fig2}
\end{figure}

Specifically, for high $\Gamma$ values, the biofilm initially forms as a compact structure, where bacterial cells grow and divide in close proximity. By contrast, as $\Gamma$ decreases, this compactness diminishes, particularly in microcolonies with a small number of cells. In Fig.\,\ref{fig1} are represented typical sequences of the evolution of biofilms for the different values of $\Gamma$ explored in this study ($\Gamma=10, 1$ and $0.1$, obtained with averaged growth rate $r\cdot\tau= 0.28, 0.028$ and $0.0027$, respectively \cite{DEL22}). For high $\Gamma$ values, the  diffusion of bacteria is dominated by the growth of individual cells, leading to compact biofilms with a subtle ellipsoidal shape (top rows). Conversely, as $\Gamma$ decreases, which implies that the diffusion time is shorter than the growth time, particle diffusion becomes more significant, yielding less compact and more circular microcolonies (bottom rows). Additionally, the dynamic evolution of the spherical tracer, indicated by the small red filled circle in Fig.\,\ref{fig1}, is shown for biomass values $m(t)$ ranging from 100 to 4000. Initially, the tracer remains stationary as cells grow and divide around it, until the biomass reaches a value of 100. Beyond this point, the tracer starts to diffuse within the colony as biomass increases. This progression is visualized in Fig.\,\ref{fig1}, where the columns from left to right correspond to increasing $m(t)$ for various $\Gamma$ values.

\begin{figure*}[!t]
	\includegraphics{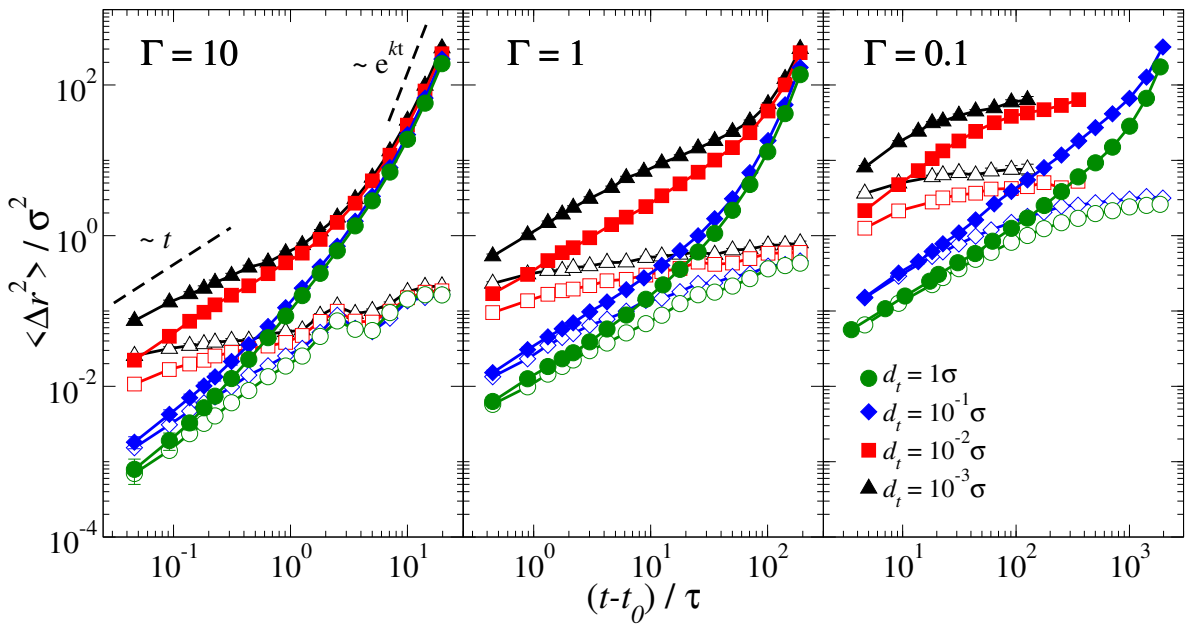}
	\caption{Mean squared displacements (MSDs) for both the radial ($\langle\Delta\textbf{r}_{\text{rad}}^2(t-t_0)\rangle$, solid symbols) and tangential ($\langle\Delta\textbf{r}_{\text{tan}}^2(t-t_0)\rangle$, empty symbols) components of the positional vector of the tracer particle projected onto the vector between the tracer position and the center of mass of the system. Four different tracer sizes were considered: $d_t=10^{-3}\sigma$ (triangles), $d_t=10^{-2}\sigma$ (squares), $d_t=10^{-1}\sigma$ (diamonds) and $d_t=1\sigma$ (circles), conducted under three distinct bacterial growth rates $\Gamma=10$ (left panel), $\Gamma=1$ (central panel), and $\Gamma=0.1$ (right panel).}\label{fig3}
\end{figure*}

\subsection{Mean Square Displacement Behaviour of a Spherical Particle in a Growing Biofilm}

As described in Section~\ref{sec_meth}, the components of the system (bacteria and tracer) lack intrinsic active motion, as their displacements arise solely from thermal fluctuations and particle-particle interactions. See Eqs.\,\ref{eq1} to \ref{eq8}. Nevertheless, the system as a whole exhibits characteristics of active motion due to the processes of bacterial growth and division. In our previous works \cite{ACE18,DEL22} we established that biomass grows exponentially with time as $m(t) \simeq \exp(rt)$, where $r$ is the average cellular growth rate. Such exponential growth is a hallmark of early biofilm formation \cite{THO06,COS95}. We hypothesize that the activity of the microcolony introduces an additional drag effect on the motion of the tracer particle, complementing the Brownian contribution. To explore this idea, we estimate the magnitude of this drag effect. Assuming that the density of the biomass remains uniform and constant throughout biofilm growth \cite{ACE18,DEL22}, the total area of the biofilm should follow the same exponential growth law as its biomass. This implies the area growth rate is proportional to the area itself: $dA/dt = kA$. Thereby, any smaller region within the biofilm centered around the colony should follow the same growth dynamics. Consequently, each point within the microcolony will notice a velocity field, with an outward drag velocity proportional to its radial distance, given by $dR/dt = k R/2$. This drag velocity, which is absent in the model’s explicit formulation, implies that the tracer particle’s mean square displacement ($\Delta \textbf{r}^2$) would grow exponentially with time ($\Delta \textbf{r}^2 \propto \exp\left(kt\right)$) and linearly with biomass ($\Delta \textbf{r}^2 \propto m(t)$), particularly dominating at long times.

To validate the described scenario, we computed the mean square displacement (MSD) of the tracer, which is defined as

\begin{equation}\label{res1}
\left\langle \Delta \textbf{r}^2 (t-t_0)\right\rangle = \left\langle (\textbf{r}(t)-\textbf{r}(t_0))^2 \right\rangle,
\end{equation}

where $\textbf{r}(t)$ indicates the position of the tracer particle at time $t$, while $t_0$ marks the time at which the computation of the MSD begins. The angular brackets represent an average over multiple trajectories. The averages include only those trajectories where the tracer particle remains within the colony at each time step. To determine whether the tracer has abandoned the colony, the best-fit ellipse describing the bacterial distribution at each time instant was computed, following the method outlined in \cite{DEL22}. If the tracer crosses the boundary of this ellipse, it is considered to have left the colony, and subsequent points in that trajectory are excluded from the averaging process. 

\begin{figure*}
	\includegraphics{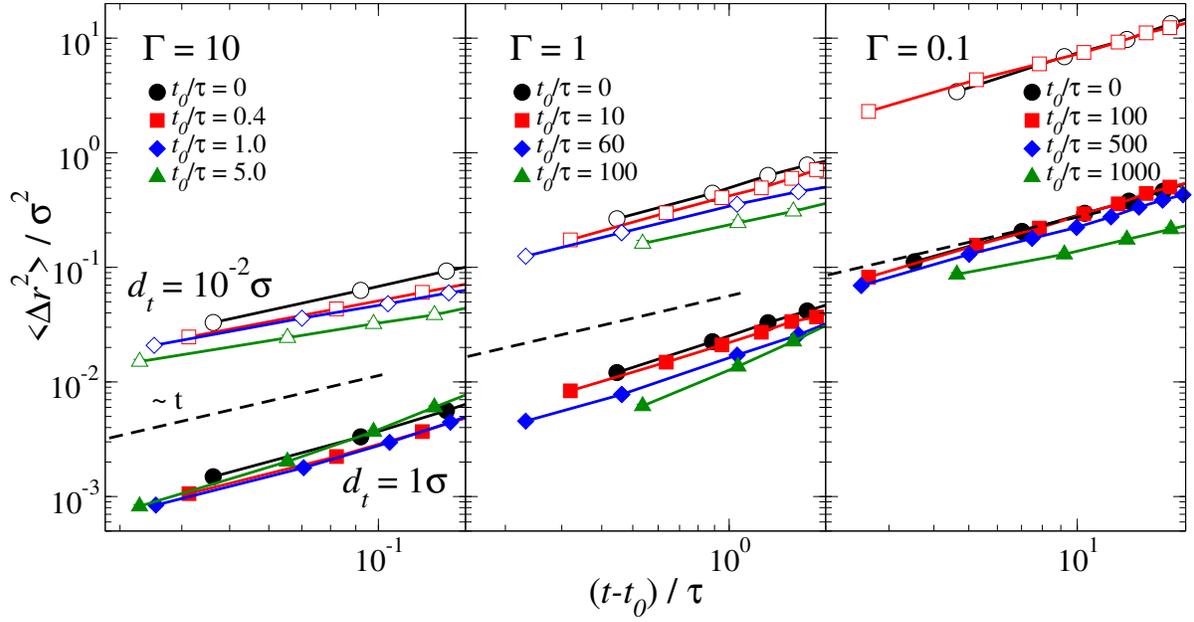}
	\caption{Short-term mean-squared displacements of tracers with sizes $d_t=10^{-2}\sigma$ (empty symbols) and $d_t=1\sigma$ (filled symbols) computed from different initial time points $t_0$ throughout the simulations. These calculations were conducted under three distinct bacterial growth scenarios: $\Gamma=10$ (left panel), $\Gamma=1$ (central panel), and $\Gamma=0.1$ (right panel).}\label{fig4}
\end{figure*}

As shown in Fig.\,\ref{fig2}, this hypothesis is validated by the MSD behavior observed in the BD simulations. The figure displays the MSD relative to the biomass for $\Gamma=10$, $1$, and $0.1$ (from top to bottom) at various tracer particle sizes ($d_t/\sigma=1, 10^{-1}, 10^{-2}$, and $10^{-3}$). The insets illustrate the time dependence of the MSD. At long times, the MSD scales linearly with biomass, indicating an exponential dependence on time, as corroborated in the insets. Notably, the long time behaviour of the MSD is independent of the tracer size, collapsing onto a universal curve for each $\Gamma$. Interestingly, this collapse also occurs at short and intermediate biomass (and time) for tracers of significantly different sizes in the case of larger values of $\Gamma$. Smaller tracers ($d_t = 10^{-2}\sigma$ and $d_t = 10^{-3}\sigma$) are less influenced by bacterial expansion and show higher MSD values. Conversely, larger tracers ($d_t = 1\sigma$ and $d_t = 10^{-1}\sigma$) are more affected by collisions with neighbouring particles, reducing their MSD. This interaction creates a size-dependent separation in the MSD values. However, the separation is less pronounced at higher $\Gamma$ values (top panel), where the biofilm is denser, and tracer movement is dominated by the colony expansion. It is also worth noting that while the biomass range ($10^2$–$10^3$) is consistent across $\Gamma$ values, the timescale for MSD calculations varies with $\Gamma$. Lower $\Gamma$ values correspond to slower biofilm growth, implying longer growth times, $t_{gr}$, which is the relevant time scale when fixing the cell diffusion time. For $\Gamma = 0.1$, smaller tracers exhibit limited MSD computation at lower biomass and time due to their tendency to escape the biofilm. This behavior will be analysed later in the Discussion section.

A deeper analysis of the long-term behavior of the MSD is provided in Fig.\,\ref{fig3}. The figure displays the radial and tangential components of the MSD, decomposed along the direction of the vector connecting the tracer's position at time $t$ and the system's center of mass $\textbf{r}_{\text{CM}}(t)$ at the same instant. Specifically, the radial (or parallel) projection of the MSD along the distance vector is expressed as
\begin{equation}
	\label{eqrad}
	\Delta\textbf{r}_{\text{rad}}(t-t_0)=\frac{\left(\textbf{r}(t)-\textbf{r}(t_0)\right)\cdot\textbf{R}(t)}{|\textbf{R}(t)|}\hat{\textbf{R}}(t),
\end{equation}
where $\textbf{R(t)}=\textbf{r}_{\text{CM}}(t)-\textbf{r}(t)$, and $\hat{\textbf{R}}(t)$ represents its corresponding unit vector. Conversely, the tangential (or perpendicular) projection is given by
\begin{eqnarray}
	\label{eqtan}
	\Delta\textbf{r}_{\text{tan}}(t-t_0)&=&[|\textbf{r}(t)-\textbf{r}(t_0)|^2\nonumber\\
	&&  - |\Delta\textbf{r}_{\text{rad}}(t-t_0)|^2]^{1/2}\hat{\textbf{R}}_{\perp}(t),
\end{eqnarray}
where $\hat{\textbf{R}}_{\perp}(t)$ is a unit vector orthogonal to $\textbf{R}(t)$, confined to the plane spanned by $\textbf{r}(t) - \textbf{r}(t_0)$ and $\textbf{R}(t)$.
From the previous equations, the radial and tangential MSD projections with respect to the system’s center of mass can be determined, denoted as $\langle \Delta\textbf{r}^2_{\text{rad}}(t-t_0) \rangle$ and $\langle \Delta\textbf{r}^2_{\text{tan}}(t-t_0) \rangle$, respectively. As in Fig.\,\ref{fig2}, the reference time $t_0$ corresponds to the moment when $m(t_0) = 100$, marking the beginning of the tracer movement.

According to Fig.\,\ref{fig3}, the radial MSD component retains a linear behavior for short and intermediate times, while an exponential growth for all tracer sizes and $\Gamma$ values is observed at long times, except in the case of the smallest tracers at $\Gamma = 0.1$ (right panel). By contrast, the tangential MSD component exhibits a subdiffusive behavior at long times, characterized by a power-law dependence with an exponent less than one. This is a signature of long-time diffusion for particles undergoing Brownian motion in complex fluids. These findings confirm that, at long times, tracer particles are radially dragged outward due to the microcolony's expansion, while Brownian motion drives the tangential diffusion. The Brownian contribution results from collisions with surrounding particles, which induces subdiffusive dynamics, as observed in other complex fluids \cite{SOK12,BIE08}. This effect is evident in the tangential MSD component $\left\langle \Delta \textbf{r}_{\text{tan}}^2 (t-t_0)\right\rangle$ but is negligible in the radial MSD, which is dominated by the microcolony expansion. A distinct case arises for small $\Gamma$ and tracer particle sizes $d_t=10^{-3}\sigma$ and $d_t=10^{-2}\sigma$: here, the reduced compactness of the biofilm allows the tracers to escape the radial drag caused by microcolony expansion, leading to subdiffusive motion dominated by Brownian diffusion.

So far we have focused our attention to the time and biomass dependence of the tracer MSD in the long-time limit. Let us now turn our attention to the short-time behavior. Figures \ref{fig2} and \ref{fig3} demonstrate that in this regime, the tracer exhibits a diffusive motion characterized by a linear relationship between the MSD and time. The diffusion coefficient, which serves as the proportionality constant, varies with the tracer particle size. This observation is further supported by Fig.\,\ref{fig4}, which shows the short-time MSD of the tracer for various initial times $t_0$ across three different values of $\Gamma$, with tracer diameters $d_t=10^{-2}\sigma$ and $d_t=1\sigma$. In this context, $t_0=0\tau$ represents the moment when $m(t)=100$, marking the beginning of the tracer's free motion. The individual panels in Fig.\,\ref{fig4} confirm that at times much shorter than the bacterial growth time $t_{\text{gr}}$, the MSD retains a linear dependence on time, indicative of Brownian motion. This linear trend is consistent across all the values of $\Gamma$, with the effective diffusion coefficient (derived from the vertical-axis intercept in the log-log plot) depending on the tracer diameter and showing a slight dependence on the initial time $t_0$. Interestingly, this $t_0$ dependence is absent for the tracer with $d_t=1\sigma$ at $\Gamma=10$ and for the smallest tracer when $\Gamma=0.1$. This will be discussed further below.

\begin{figure*}[!t]
	\includegraphics[width =\columnwidth]{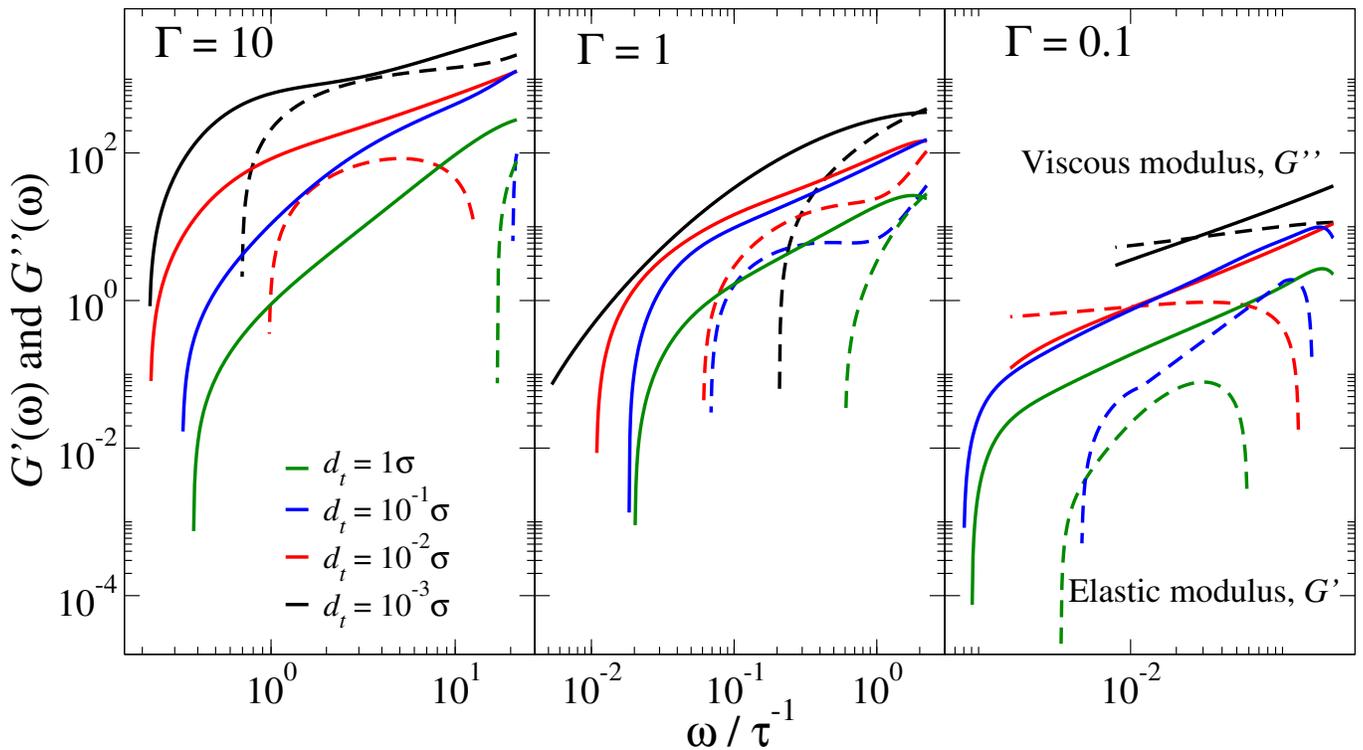}
	\caption{Viscous ($G''$, solid lines) and elastic ($G'$, dashed lines) moduli calculated from the MSD of tracers with sizes  $d_t=10^{-3}\sigma$ (black lines),  $d_t=10^{-2}\sigma$ (red lines),  $d_t=10^{-1}\sigma$ (blue lines) and  $d_t=1\sigma$ (green lines) in microcolonies with values of $\Gamma=10$ (left), $\Gamma=1$ (middle) and $\Gamma=0.1$.}\label{fig7}
\end{figure*}	

\begin{figure*}
	\includegraphics{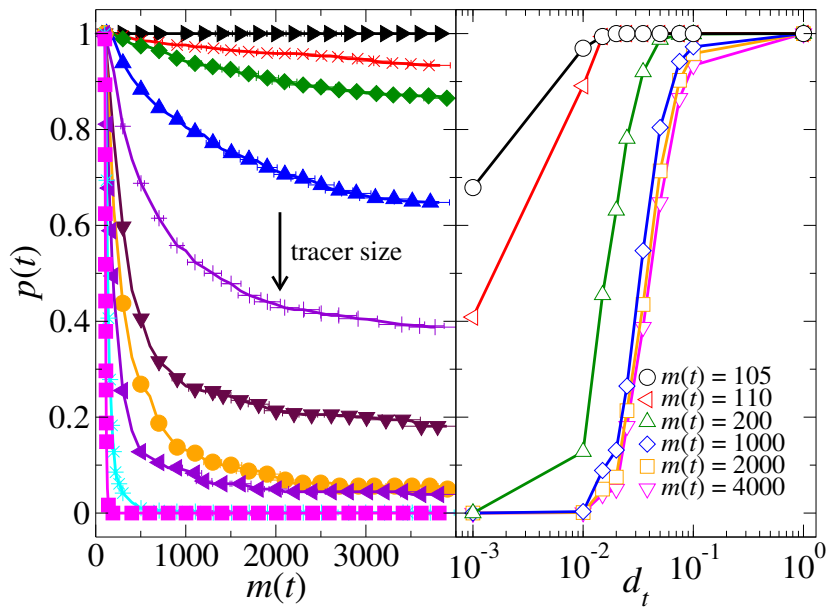}
	\caption{probability of a tracer staying within the colony, $p(t)$, within a growing bacterial colony with a growth rate $\Gamma=0.1$. Left panel: $p(t)$ is depicted for different tracer sizes (from top to bottom, $d_t/\sigma=1, 10^{-1}, 7.5\cdot10^{-2}, 5\cdot10^{-2}, 3.5\cdot10^{-2}, 2.5\cdot10^{-2}, 2\cdot10^{-2}, 1.5\cdot10^{-2}, 10^{-2}, 10^{-3}$) as a function of the system biomass. The right panel displays $p(t)$ for a specific instantaneous biomass, with variations based on the tracer size.}
	\label{fig5}
\end{figure*}
\subsection{Scaling Effects of Tracer Size and $\Gamma$ on the Rheology of Bacterial Colonies}

As outlined in the Introduction section, tracer diffusion within a complex medium has been recognized as an effective approach for determining rheological properties via microreology techniques. Consequently, in passive MR, the methodology introduced by Manson and colleagues \cite{MAN95,MAN00} allows for the estimation of the viscoelastic characteristics of the host material based on the computed MSD. Following Manson’s framework \cite{MAN00}, the shear modulus can be expressed as

\begin{equation}
\label{eq14}
\left|G^{*}\left(\omega\right)\right| = \frac{k_{\text{B}}T}{\pi \left(d_t/2\right) \langle\Delta r^2_{t}(1/\omega)\rangle\Gamma\left[1+\alpha\left(\omega\right)\right]},
\end{equation}
\noindent where $\omega=1/t$ is the time frequency, $\alpha\left(\omega\right) \equiv \left(d\ln\langle\Delta r^2_{t}(t)\rangle/d\ln\left(t\right)\right)|_{t=1/\omega}$ is the local exponent of the MSD, and $\Gamma[1+\alpha(\omega)]$ is the gamma function (not to be confused with parameter $\Gamma$). Correspondingly, the elastic or loss ($G'$), and viscous or storage moduli ($G''$) can be determined using the following equations:

\begin{align}
G'(\omega) &= \left|G^{*}\left(\omega\right)\right|\cos \left(\frac{\pi\alpha(\omega)}{2}\right),
\label{eq15}\\
G''(\omega) &= \left|G^{*}\left(\omega\right)\right|\sin \left(\frac{\pi\alpha(\omega)}{2}\right).
\label{eq16}
\end{align}

We highlight that the Mason’s approximation, as used here, assumes linear response and equilibrium conditions. In the context of a dynamically evolving biofilm, these assumptions may not fully hold, especially at extreme frequencies, as previously reported by Mason et al. \cite{MAN95,MAN00}. As such, our microrheological analysis should be interpreted as a qualitative rather than strictly quantitative description in the absence of experimental validation. Future work will focus on benchmarking our computational predictions against relevant experimental microrheological measurements in comparable two-dimensional biological systems. 

Figure \ref{fig7} depicts the viscoelastic moduli for three values of $\Gamma$ as a function of tracer size. The results reveal a pronounced dependence of the moduli on tracer diameter. Specifically, for larger $\Gamma$, corresponding to more compact biofilms, the system exhibits a consistently higher loss modulus than storage modulus ($G'' > G'$), independent of tracer size or frequency. This suggests that the biofilm environment is predominantly perceived as a viscous medium, an effect that becomes more apparent for larger tracers. In the case of $\Gamma = 10$, this trend is particularly evident at both high and low-frequency limits, with the latter showing complete dominance of $G''$ for larger tracers, effectively reducing $G'$ to near zero. This observation aligns with the long-time MSD analysis, which demonstrated that the tracers are transported outward by colony growth, behaving similarly to particles in a flowing viscous medium. At high frequencies, the trend is consistent with Fig.~\ref{fig4}, where tracers initially undergo diffusion before encountering cell-induced confinement. Interestingly, at intermediate frequencies and for smaller tracers, $G'$ and $G''$ become comparable, indicating that the tracers experience a subdiffusive regime where the internal structure of the colony partially restricts motion and imparts a viscoelastic character to the environment. Conversely, larger tracers are excluded from the accessible free volume of local bacterial clusters, and their motion is instead dominated by collective drag forces, leading to a more viscous-like mechanical response where $G'$ approaches $G''$.
 
At the intermediate value of $\Gamma = 1$, we observe a predominantly viscous response at low frequencies ($G'' >> G'$), which gradually evolves into a more viscoelastic behaviour at intermediate frequencies. At higher frequencies (shorter timescales), the storage and loss moduli become comparable, reflecting a transitional regime in which diffusion and colony expansion are balanced. During this timescale window, the tracer exhibits subdiffusive dynamics, wherein its motion is hindered by surrounding particles, though not entirely restricted. Consequently, the mechanical response at short timescales becomes mixed, with comparable elastic and viscous contributions ($G'' \sim G'$). Finally, for the lowest value of $\Gamma$, the results exhibit particularly intriguing behavior. Specifically, for $\Gamma = 0.1$ (right panel of Fig.~\ref{fig7}), while larger tracers maintain the previously described relationship between $G''$, $G'$, and $\omega$, smaller tracers ($d_t=10^{-3}\sigma$ and $d_t=10^{-2}\sigma$) show significant deviations. At low frequencies, the elastic modulus exceeds the viscous modulus ($G' > G''$), with a crossover at intermediate $\omega$ to a regime dominated by viscosity. This behavior aligns with the MSD trends in Fig.\,\ref{fig4}, where long-time subdiffusion suggests that the tracer experiences the microcolony as an elastic structure, whereas at short times (high frequencies), a diffusive, viscosity-dominated response emerges in both the viscoelastic moduli and the MSD. For larger tracers ($d_t = 10^{-1}\sigma$ and $d_t = 1\sigma$), the surrounding medium is perceived as predominantly viscous at high frequencies. This corresponds to a standard diffusive behavior at short times, as shown in Fig.\,\ref{fig4}. At intermediate frequencies, the tracer transitions into a subdiffusive regime where local particle interactions become more significant. In this regime, the system exhibits viscoelastic characteristics, with $G’$ approaching $G''$. As time grows, this behavior shifts back toward a viscous-dominated response ($G'' >> G'$) due to the persistent influence dragged motion imposed by the growing colony at long times (low frequencies).

\subsection{Overview of Transport Dynamics of Probes in the Early Stages of Biofilm Formation}
We can now characterize the motion of tracers within an early-stage biofilm and the structural information this motion reveals. At short times, tracers primarily diffuse through the voids and cavities formed by bacteria within the biofilm. This diffusive behavior is determined by the medium, typically the extracellular polymeric substance (EPS) matrix in real biofilms, where the matrix's viscosity controls the tracer’s diffusion coefficient. Over time, bacterial growth and division increase the pressure within the colony, leading to compaction and reduced connectivity between cavities. This results in a gradual decrease in the diffusion coefficient, as illustrated in Fig.\,\ref{fig4}, showing how tracer diffusion reflects biofilm ageing. Notably, this evolution is absent in two cases: (1) for the largest tracer size ($d_t=1\sigma$) at $\Gamma=10$, where the colony is compact from the outset, preventing significant diffusion, and (2) for the smallest tracer size ($d_t=10^{-2}\sigma$) at $\Gamma=0.1$, where the colony remains open, allowing free tracer diffusion.

At later times, as in other complex fluids, collisions with bacteria begin to dominate the dynamics, altering the regime of tracer motion. Typically, in complex fluids, such collisions lead to subdiffusive behavior, characterized by an MSD that scales with time as a power law with an exponent less than one. This phenomenon is evident in the tangential component of the MSD but not in the radial component (see Fig.\,\ref{fig3}). Along the radial direction, the radial expansion of the microcolony induces a drag force on the tracer, causing the radial MSD to exhibit an exponential time dependence. However, for smaller tracers ($d_t = 10^{-2}\sigma$ and $10^{-3}\sigma$) at low $\Gamma$, this exponential trend is absent, being replaced by a subdiffusive behavior. This suggests that for smaller tracers and low $\Gamma$, motion within the biofilm is not affected by the drag induced by microcolony’s expansion, as it is observed in other limits. Interestingly, at the limit of low $\Gamma$, small tracer size, and low $\omega$ (i.e., long timescales), this observation is consistent with the MR analysis presented in Fig.\ref{fig7} (right panel). Under these conditions, the particle begins to perceive the biofilm as a more elastic than viscous medium, with its motion increasingly influenced by the local microcolony structure. However, we emphasize that this particular behavior is confined to these specific conditions, as the colony microstructure does not appear to significantly affect the probe dynamics in other regimes of $\Gamma$ and tracer sizes at long times (as shown in the left and central panels of Fig.\ref{fig7}).

\subsection{Probability of Escape of a Tracer Particle from a Growing Biofilm}

The observations raise the question of whether the tracer’s combined motions allow it to escape the microcolony from its initial central position. The general conclusion is, for almost all $\Gamma$ values, that escape does not occur. The exception would appear for low $\Gamma$ values and small tracer particles. Furthermore, simulations reveal that, regardless of the tracer’s diameter or $\Gamma$ values of $10$ and $1$, no tracer crosses the microcolony boundary. Escape is defined as the tracer surpassing the ellipse that optimally fits the bacterial distribution within the colony. This behavior is explained by the dominance of drag forces arising from the colony’s growth and swelling, which push the tracer outward. However, the boundary of the colony expands outward at the same time, remaining beyond the tracer’s reach despite its diffusive motion.

For low values of $\Gamma$, such as $\Gamma = 0.1$, this behavior warrants further clarification. Figure \ref{fig5} presents, for $\Gamma=0.1$ the probability of a tracer staying within the colony, $p(t)$, for various diameters $d_t$ as the biofilm evolves, characterized by the biomass $m(t)$ (see left panel). This probability is computed as the fraction of trajectories in which the tracer remains within the colony, relative to the total number of trajectories under the same $d_t$ and $m(t)$. The figure reveals that for $d_t = 1\sigma$ (black right triangles), the tracer remains confined within the colony throughout the observed time frame, even though the colony is not particularly compact at this diameter, where diffusion might reasonably facilitate escape. For smaller tracers, however, the probability of staying within the colony diminishes with both decreasing diameter and increasing biomass. Thus, for $d_t \leq 10^{-2}\sigma$, the probability of residence in the microcolony falls to zero almost immediately. By contrast, tracers with $d_t \geq 10^{-1}\sigma$ are consistently retained within the colony, maintaining a near-perfect probability of survival inside the colony for all biomass values. 

Finally, we evaluated how the instantaneous biomass influences the probability of staying within the colony of tracers depending on their size, as illustrated in the right panel of Fig.\,\ref{fig5}. During the initial stages of the colony expansion ($m(t) \leq 110$), only tracers with diameters $d_t \geq 10^{-2}\sigma$ are likely to remain within the colony. By contrast, smaller tracers ($d_t < 10^{-2}\sigma$) are more likely to escape due to their ability to diffuse through the spaces between host particles formed in the early microcolony stages. As the biomass increases, the range of tracer sizes able to persist in the colony is reduced to particles with $d_t \ge  10^{-1}\sigma$. This panel shows that for $\Gamma=0.1$, smaller diameter particles will escape from the microcolony as the biomass increases.

\section{Final remarks and conclusions}

In summary, based on the results discussed in the previous sections, it can be inferred that particle transport within growing biofilms follows two fundamental scenarios. At very short times, the tracer particle undergoes a diffusive Brownian motion. This motion is caused by thermal agitation and causes the particle to diffuse through the intercellular medium of the biofilm. On the other hand, at very long times the particle experiences a drag motion due to the growth of the cell colony. In this scenario the particle is mainly dragged outwards, with a radial MSD that depends exponentially on time and linearly on biomass. At intermediate times a subdiffusive intermediate behaviour is observed (power-law dependence of the MSD with time, with an exponent less than one). This subdiffusive regime is evident in the tangential component of the MSD projection onto the vector between the tracer and the center of mass of the system, but in most situations is not clear in the radial component (the dominant component of the motion). We stress that this description is applicable to the majority of the cases analyzed in this study, where $\Gamma$ varies between 0.1 and 10. This range results in significant changes in the colony's compactness, with particle sizes spanning from those comparable to the cell diameters to those $10^{3}$ times smaller. Specifically, this encompasses a size range from micrometers to nanometers, where antibiotic molecules, proteins, viruses, and nanoparticles may be found, among others. However, very small particles diffusing into loosely packed cell colonies may not exhibit the same general behavior.

These general scenarios are complemented through passive microrheological analysis by evaluating the biofilm's elastic and viscous characteristics. Simulated microcolonies exhibit fluid-like behavior at both low and high frequencies, while demonstrating viscoelasticity at intermediate frequencies. Interestingly, a key exception emerges: in colonies with low $\Gamma$, small tracers experience a dominant elastic response at short frequencies. It is crucial to recognize that our study isolates the impact of growing cells on $G'$ and $G''$, omitting potential influences from other biofilm components, particularly the EPS matrix, which may significantly alter these parameters. Nevertheless, focusing exclusively on the active contributions from cell growth and division, we observe a particularly noteworthy mechanical response.

The characteristics observed in our simulations and discussed in this paper have also been seen in various types of growing cell communities. For example, Sinha and co-workers have reported the existence of superdiffusive motion at long times in tracer particles within growing multicellular spheroids of cancer cells \cite{SUM23}. Similar results in animal cell and tissues have been reached by other authors \cite{WEI06,KRA21}. Although these works label this motion as superdiffusive, the MSD exhibits a time dependence of the form $\text{MSD}\sim t^{\alpha}$, with $\alpha > 1$. 

However, based on their findings, it remains unclear whether the underlying mechanism in those systems is the same as the one we have modeled. To verify this, it would be useful to investigate whether the MSD’s time dependence is truly exponential. As for bacterial biofilms, published results appear less conclusive, as experimental findings report the occurrence of both subdiffusive and superdiffusive scenarios \cite{ROG08,VAS22,WAI23}. These studies primarily focus on mature biofilms, and to the best of our knowledge, research on the diffusion of tracer particles in early-stage biofilms, like those modeled in this study, remains limited.
 
We highlight, however, that this work does not examine the diffusion of the tracer particle in the later phases of biofilm expansion, where constraints on cellular growth, such as nutrient depletion, become increasingly relevant. At these advanced stages, the biofilm can grow to the point where its boundary is significantly distant from the tracer particle, reducing the influence of outward advection as the primary transport mechanism. Consequently, the biofilm evolves into a more intricate three-dimensional structure. Understanding how this transition in dimensionality and structural complexity impacts tracer particle diffusion and the microrheological properties of the surrounding microcolony is essential. We expect that future research within our group will explore these aspects in greater detail.

\begin{acknowledgments} This work was supported by the Spanish Ministerio de Ciencia e Innovaci\'on and FEDER (project no. PID2021-126121NB-I00), and the Consejería de Transformaci\'on Econ\'omica, Industria, Conocimiento y Universidades de la Junta de Andaluca/FEDER (project no. P20-0081). F.A.G.D. was funded by the NextGenerationEU program of the European Union, the Plan de Recuperaci\'on, Transformaci\'on y Resiliencia, and the Ministerio de Universidades, as part of the ``Maria Zambrano'' grants for the requalification of the Spanish university system 2021-2023 called by the Pablo de Olavide University. A R.-R. also acknowledges financial support from Grant No. PID2021-126348NB-I00 funded by MCIN/AEI/10.13039/501100011033.  The authors are thankful to the C3UPO of the Pablo de Olavide University for the support with HPC facilities.
\end{acknowledgments}


\providecommand{\noopsort}[1]{}\providecommand{\singleletter}[1]{#1}%

\end{document}